\documentclass[aps,pra,amsmath,amssymb,twocolumn,showpacs,groupedaddress]{revtex4-1}

\usepackage{graphicx}
\usepackage{amsbsy}
\usepackage{amssymb}
\usepackage{amsfonts}
\usepackage{amsmath}
\usepackage{bm}
\usepackage{mathrsfs}
\usepackage{color}
\usepackage{hyperref}

\newcommand{\ab}{{\hat a}}
\newcommand{\ac}{{\hat a}^\dagger}

\begin{document}

\title{Spontaneous symmetry breaking in a quadratically driven nonlinear photonic lattice}

\author{Vincenzo Savona}
\email{vincenzo.savona@epfl.ch}
\affiliation{Institute of Physics, Ecole Polytechnique F\'{e}d\'{e}rale de Lausanne (EPFL), CH-1015 Lausanne, Switzerland}

\date{\today}

\begin{abstract}
We investigate the occurrence of a phase transition, characterized by the spontaneous breaking of a discrete symmetry, in a driven-dissipative Bose-Hubbard lattice in the presence of two-photon coherent driving. The driving term does not lift the original $U(1)$ symmetry completely and a discrete $\mathbb{Z}_2$ symmetry is left. When driving the bottom of the Bose-Hubbard band, a mean-field analysis of the steady state reveals a second-order transition from a symmetric phase to a quasi-coherent state with a finite expectation value of the Bose field. For larger driving frequency, the phase diagram shows a third region, where both phases are stable and the transition becomes of first order.
\end{abstract}

\maketitle

\section{Introduction}
Critical phenomena in driven-dissipative many-body quantum systems are emerging as a major field of study. Intense efforts are being devoted to investigate new classes of phase transitions, and in view of the realization of dissipative quantum simulators using optical or superconducting circuit platforms \cite{Carusotto2013,Hartmann2016,LeHur2016,Noh2017}. These studies led very recently to the first experimental evidence of dissipative phase transitions in systems of ultra cold atoms \cite{Baumann2010,Baumann2011,Brennecke2013,Dimer2007}, superconducting circuits \cite{Carmichael2015,Fink2017a,Fitzpatrick2017}, and semiconductors \cite{Rodriguez2017,Fink2017}.

The photonic nature of most of these platforms has stimulated in particular the study of the driven-dissipative Bose-Hubbard model, which is naturally realized by an array of optical resonators in the presence of a Kerr nonlinearity. Several theoretical studies have investigated the occurrence of phase transitions under various settings \cite{Biella2017,Biondi2016,Cao2016,Casteels2017,Finazzi2015,Foss-Feig2017,LeBoite2014,LeBoite2013,Casteels2017a,Lebreuilly2017,Wilson2016}. The optical bistability inherited by the single Kerr resonator, in particular, has been shown to give rise to a critical behavior when the resonators are linearly driven by a coherent resonant field \cite{Biondi2016,Foss-Feig2017,LeBoite2014,LeBoite2013,Wilson2016}. 

One outstanding question in this domain is, whether a driven-dissipative system can reproduce the critical behavior of the closed Bose-Hubbard model, where the spontaneous breaking of the $U(1)$ symmetry of the Bose-field results in a transition from the Mott to a long-range coherent phase. In a driven-dissipative system, the resonant driving field results in the pinning of the local phase of the resonators, thus lifting from start the $U(1)$ symmetry of the underlying Bose-Hubbard system and hindering the phase transition. Very recently, it was shown that the Mott-like physics can be recovered via an incoherent, non-Markovian driving term with narrow-band noise spectrum \cite{Biella2017,Lebreuilly2017}.

In a different context, the system of a single Kerr resonator in the presence of a {\em two-photon} coherent driving term has been extensively studied, both theoretically \cite{Bartolo2016,Bartolo2017,Goto2016,Goto2016a,Minganti2016,Nigg2017,Puri2017a,Puri2017} and experimentally \cite{Leghtas2014}. This quadratically driven Kerr resonator naturally realizes a steady state which is a statistical mixture of two Schr\"odinger's cat states, each being a linear superposition of coherent states with opposite displacements. This system is an ideal framework for fundamental studies on decoherence mechanisms in macroscopic nonclassical states, and is experimentally viable in particular with circuit-QED systems \cite{Leghtas2014}. In addition, a network of coupled, quadratically driven Kerr resonators, leveraging on the occurrence of photonic Schr\"odinger's cat states, has been proposed as a physical realization of a quantum annealer \cite{Goto2016,Goto2016a,Nigg2017,Puri2017a,Puri2017}, with potentially disruptive impact on quantum information technologies \cite{Devoret2014,Gilchrist2004,Ourjoumtsev2006,Vlastakis2013,Leghtas2014,Ofek2016,Wang2016}.

The occurrence of a photonic Schr\"odinger's cat state can be traced back to the symmetry of the system in the presence of the two-photon driving term. This term sets the complex phase of the {\em square} of the cavity field. Then, the initial $U(1)$ symmetry of the system is only partially lifted, and a discrete $\mathbb{Z}_2$ symmetry is left, corresponding to solutions with opposite values of the complex field amplitude. This feature represents a major difference with respect to the linearly driven case, where the driving field lifts the $U(1)$ symmetry completely. The question then naturally arises, whether in the quadratically driven lattice a spontaneous symmetry breaking can occur, giving rise to a phase transition between a $\mathbb{Z}_2$-symmetric phase and a coherent phase with nonzero expectation value of the Bose field.

In this paper, we study the occurrence of this spontaneous symmetry breaking within a mean-field description of the quadratically driven, dissipative Bose-Hubbard lattice. The occurrence of a phase transition between a symmetric and a broken-symmetry phase is evidenced by three independent approaches: a stability analysis of the excitations characterizing the symmetric phase, the analytical calculation of the steady state, and the simulation of the mean-field dynamics described by the master equation. In addition to these two phases, depending on the frequency detuning of the two-photon driving field, the phase diagram may display a third region where both solutions are stable and their occurrence depends on the specific system dynamics.

The article is organized as follows. In Section II we derive the mean-field theory of the quadratically driven array of coupled Kerr resonators. Section III presents the results of the numerical analysis. In Section IV we discuss the implications of the results and the possible physical implementations. Section V contains the conclusions and the outlook of this paper.

\section{Theory}
The full Hamiltonian of the quadratically driven Bose-Hubbard model is 
\begin{equation}
\mathcal{\hat H} = \sum_j\hat h_j+\frac{J}{z}\sum_{\langle j,k\rangle}\left(\ac_j\ab_k+\ac_k\ab_j\right)\,,
\label{H}
\end{equation}
where $\ac_j$ and $\ab_j$ are the bosonic creation and annihilation operators for the $j$-th site, $J$ is the hopping strength, $z$ is the coordination number of each site, and the second sum runs over pairs of neighboring sites in the lattice. Here, $\hat h_j$ is the single-site Hamiltonian which, in the rotating frame of the resonator, is expressed as
\begin{equation}
\hat h_j=-\Delta\ac_j\ab_j+\frac{U}{2}\ac_j\ac_j\ab_j\ab_j+\frac{G}{2}\ac_j\ac_j+\frac{G^*}{2}\ab_j\ab_j\,.
\label{hj}
\end{equation}
In this expression, $U$ is the strength of the Kerr nonlinearity, $G$ is the amplitude of the two-photon driving. In the rotating frame, $\Delta=\omega_2/2-\omega_c$, where $\omega_2$ is the frequency of the two-photon driving and $\omega_c$ is the resonator frequency (we set $\hbar=1$). 

\begin{figure}[ht!]
\includegraphics[width=0.50 \textwidth]{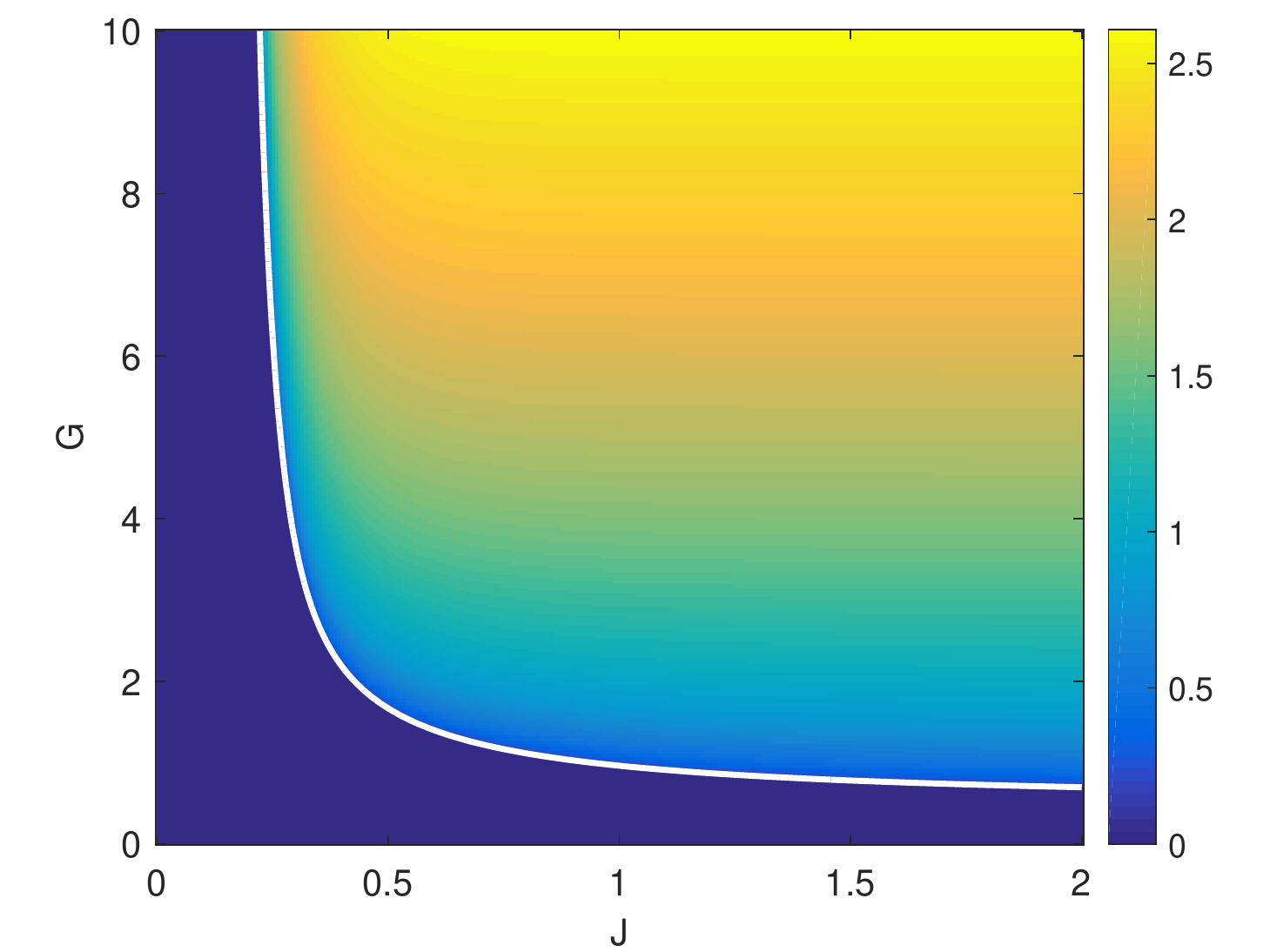}
\caption{\label{fig1} White line: Contour separating the regions where $\max(\mathrm{Im}(\omega_k))|_k<0$ (left) and $\max(\mathrm{Im}(\omega_k))|_k>0$ (right) on the $(J,G)$-plane. Color plot: the order parameter $|\langle\ab\rangle|$ computed self-consistently, at steady-state, from the analytical solution of Ref. \onlinecite{Bartolo2016}. Parameters: $\Delta=-J$, $U=\eta=1$.}
\end{figure}

The mean-field approximation consists in assuming a fully factorized form for the density operator of the system $\hat\rho_{sys}=\bigotimes_j\hat\rho_j$. Dropping the index $j$ from the notation, the single-site density operator $\hat\rho$ then obeys the master equation
\begin{equation}
\frac{d\hat\rho}{dt}=\mathcal{L}\hat\rho=-i[{\mathcal{\hat H}_{MF}},\hat\rho]+\frac{\kappa}{2}\mathcal{D}(\ab)\hat\rho+\frac{\eta}{2}\mathcal{D}(\ab^2)\hat\rho\,,
\label{master}
\end{equation}
where the dissipation super-operators defined as $\mathcal{D}(\hat K)\hat\rho=-\{\hat K^{\dagger}\hat K,\hat\rho\}+2\hat K\hat\rho \hat K^{\dagger}$ model losses into the environment within the Born-Markov approximation. In our model, we assume for each resonator both one- and two-photon loss processes \cite{Bartolo2016,Bartolo2017,Minganti2016}, with rates $\kappa$ and $\eta$ respectively. Two-photon losses are not determinant to the physics described below \cite{Bartolo2016}. Their inclusion is however natural as, in an open system, a two-photon input channel will in general operate also as an output channel to the environment. The corresponding mean-field Hamiltonian is 
\begin{eqnarray}
\mathcal{\hat H}_{MF}&=&-\Delta\ac\ab+\frac{U}{2}\ac\ac\ab\ab+\frac{G}{2}\ac\ac+\frac{G^*}{2}\ab\ab\nonumber\\
&+&J(\langle\ab\rangle\ac+\langle\ab\rangle^*\ab)\,,
\label{Hmf}
\end{eqnarray}
where the mean field amplitude is defined self-consistently as $\langle\ab\rangle=\mathrm{Tr}[\ab\hat\rho]$. In what follows, all energies and time are expressed in units of $\kappa$ and $\kappa^{-1}$ respectively, and we will assume $U=\eta=1$.

The symmetric steady-state solution $\hat\rho_s$ with $\langle\ab\rangle=0$ is always admitted by the mean-field model. It coincides with the solution of the corresponding model of the single Kerr resonator with two-photon driving, which has been extensively discussed in the literature \cite{Bartolo2016,Bartolo2017,Goto2016,Goto2016a,Minganti2016,Nigg2017,Puri2017a,Puri2017}. The steady state of such a system is a statistical mixture which, in the limit of large driving $G$, is dominated by two Schr\"odinger's cat states of opposite parity $|{\cal C}^\pm_\alpha\rangle=(|\alpha\rangle\pm|-\alpha\rangle)/\sqrt{2(1\pm e^{-2|\alpha|^2})}$, where $|\alpha\rangle$ is a coherent state. For this steady state, $\langle\ab\rangle=0$ rigorously holds as a consequence of the $\mathbb{Z}_2$ symmetry. The mixture of two cat states is equivalent to a mixture of the two coherent states $|\pm\alpha\rangle$ with opposite field displacements. In this sense, the spontaneous breaking of the $\mathbb{Z}_2$ symmetry should be understood as the occurrence of a state closer to one of the two pure states $|\pm\alpha\rangle$.

To investigate the existence of a phase with a spontaneously broken symmetry, we first study the stability of the symmetric steady-state solution obtained by setting $J=0$ and solving Eq. (\ref{master}) for $d\hat\rho_s/dt=0$. An equation for the excitations $\delta\hat\rho_j$ can be derived starting from the factorized ansatz $\hat\rho=\bigotimes_j(\hat\rho_s+\delta\hat\rho_j)$ and carrying out a linear expansion of the master equation around the symmetric steady-state solution $\hat\rho_s$ \cite{LeBoite2014,Schiro2016}. The resulting equation reads
\begin{equation}
-i\omega_k\delta\hat\rho_k=\mathcal{L}\delta\hat\rho_k-t_k\left[i\mathrm{Tr}(\ab\delta\hat\rho_k)[\ac,\hat\rho_s] + \mathrm{h.c.}\right]\,,
\label{exc}
\end{equation}
where the Liouvillian super-operator $\mathcal{L}$ is the one defined in (\ref{master}) for $\langle\ab\rangle=0$, and we have introduced the momentum representation through $\delta\hat\rho_j=\sum_k\delta\hat\rho_ke^{i(kj-\omega_kt)}$. The quantity $t_k=-J\cos(k)$ is the dispersion of the corresponding lattice of harmonic oscillators (i.e. $U=0$), and we assumed a one-dimensional lattice model with $z=2$ for simplicity. If the symmetric solution is stable, the eigenvalues $\omega_k$ obtained from Eq. (\ref{exc}) will all have negative imaginary part, corresponding to damped excitations. A positive value of $\mathrm{Im}(\omega_k)$ on the other hand is a signature of the possible existence of different stable solutions. We therefore consider the sign of the quantity $\max(\mathrm{Im}(\omega_k))|_k$ as an indicator of the stability of the symmetric solution.

\section{The mean-field phase diagram}
In the first part of this analysis, we set $\Delta=-J$. This choice corresponds to tuning the quadratic driving field in resonance with the bottom of the band of the Bose lattice at frequency $t_{k=0}$, and is adopted so to avoid the bistable behavior that may occur at higher detuning \cite{Drummond1980}. The main result of the present paper is summarized in Fig. \ref{fig1}, where the white contour line on the $(J,G)$ plane separates the regions where the symmetric solution is stable (left) and unstable (right). The stability analysis of excitations therefore provides a candidate phase diagram. In order to gain deeper insight, we compute the steady-state solution by solving self-consistently the mean-field master equation (\ref{master}) for $d\hat\rho/dt=0$. To this purpose, we adopt the analytical solution, in terms of Gauss hypergeometric functions, that has been recently derived for the Kerr oscillator in the presence of both linear and quadratic driving terms \cite{Bartolo2016,Minganti2016}. This ensures that the true steady state is found, as this system can display a very slow dynamics characterized by long-lived metastable states \cite{Minganti2016}. The order parameter $|\langle\ab\rangle|$ resulting from the mean-field calculation is diplayed as a color plot in Fig. \ref{fig1}. The plot clearly shows a region where the $\mathbb{Z}_2$ symmetry is spontaneously broken and the order parameter takes a finite value. This region coincides, up to numerical accuracy, with the instability region previously found. 

\begin{figure}[ht!]
\includegraphics[width=0.5 \textwidth]{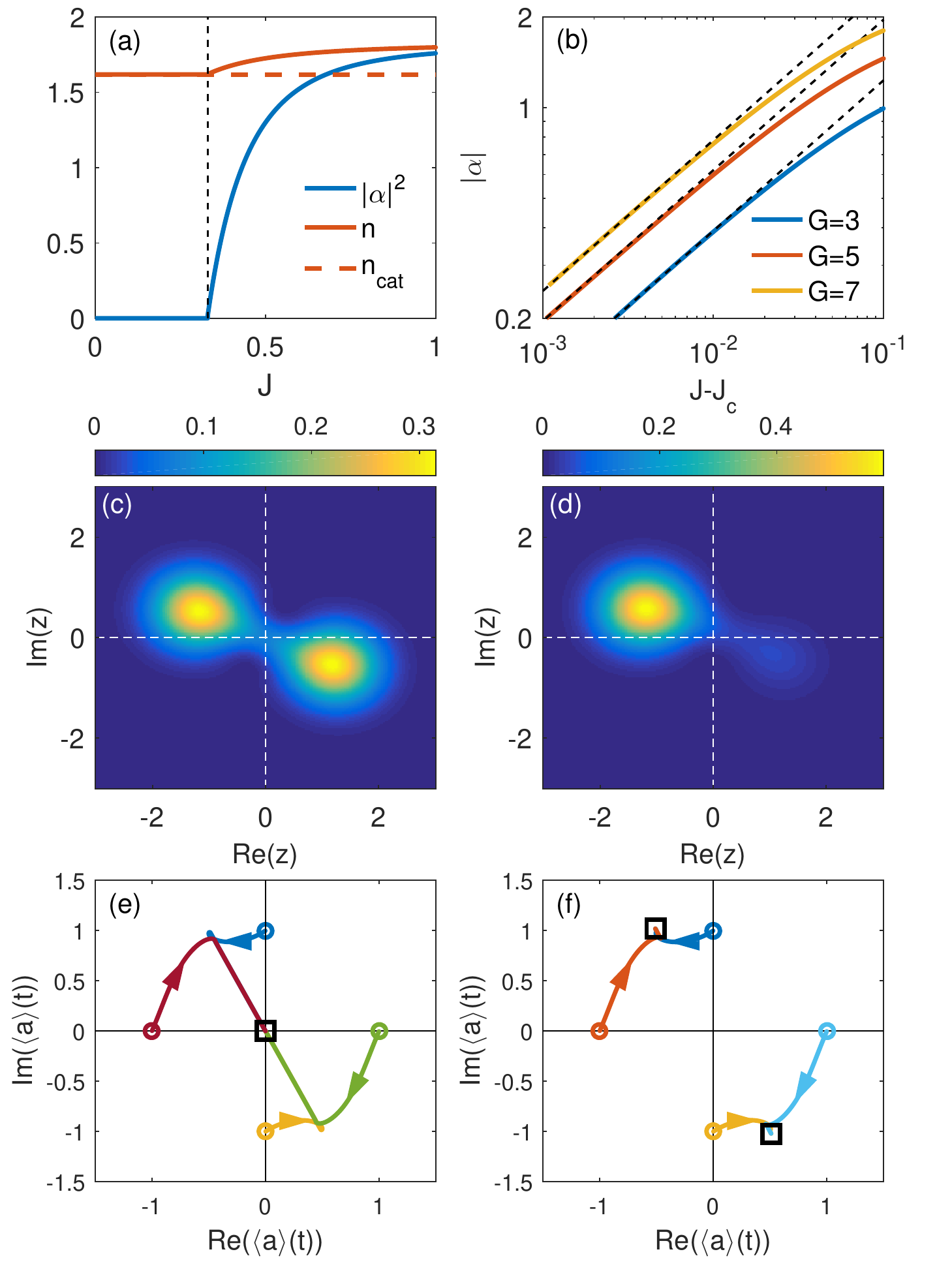}
\caption{\label{fig2} (a) The quantities $|\langle\ab\rangle|^2$ and $n=\langle\hat a^\dagger\hat a\rangle$ plotted as a function of $J$ for $G=3$. The horizontal dashed line denotes the symmetric solution $n_s$ corresponding to $|\langle\ab\rangle|=0$, and the vertical dashed line marks the fitted critical value $J_c=0.3305$. (b) Double-logarithmic plot of $|\langle\ab\rangle|$ as a function of $J-J_c$ for three values of $G$. The dashed lines denote the critical behavior $|\langle\ab\rangle|\propto|J-J_c|^{1/2}$ obtained by fitting the data close to the critical point $J=J_c$. (c) and (d) Color plot of the Wigner function $W(z)$ computed for $G=3$ and for values of $J$ corresponding, respectively, to the symmetric ($J=0.25$) and broken symmetry ($J=0.5$) phases. (e) and (f) Trajectories on the $(\mathrm{Re}(\langle\ab\rangle),\mathrm{Im}(\langle\ab\rangle))$-plane as computed for $G=3$ and, respectively, $J=0.25$ (a), $J=0.5$ (b). Different trajectories correspond to an initial coherent state $|\alpha_0\rangle$ for different values of $\alpha_0$, and the arrows indicate the direction of time. Circles denote the initial coherent states, while the squares mark the fixed points reached at steady state. Parameters: $\Delta=-J$, $U=\eta=1$.}
\end{figure}

Fig. \ref{fig2}(a) shows the quantities $|\langle\ab\rangle|^2$ and $n=\langle\ac\ab\rangle$ plotted as a function of $J$ for $G=3$. Below the critical value $J_c$, the symmetric phase is incompressible, with the average occupation $n_s$ coinciding with that of the single-site Kerr model. Above $J_c$ the occupation increases as a result of the added coherent contribution $|\langle\ab\rangle|^2$ from the order parameter. For large $J$ the state approaches a pure coherent state and the two quantities coincide in this limit. To assess the nature of this phase transition, we carry out a power-law fit according to $|\langle\ab\rangle|\propto|J-J_c|^\beta$, for data in the vicinity of $J=J_c$, the results of which are reported in Fig. \ref{fig2}(b) for $G=3,5,7$. For these -- and for all the values of $G$ considered in this analysis -- the fit results in $\beta\approx0.5$, indicating a second order phase transition with a critical exponent $\beta=1/2$ as expected in a mean-field analysis. Figs. \ref{fig2}(c) and (d) show the Wigner function $W(z)$ \cite{Bartolo2016} of the steady state, as computed for $G=3$ and, respectively, for $J=0.25$ and $J=0.5$, corresponding to the two different phases. The first case coincides with the solution of the single-site Kerr model \cite{Bartolo2016,Bartolo2017,Minganti2016}. The second plot depicts one of the two possible quasi-coherent states resulting from the spontaneous symmetry breaking. As an additional witness of the nature of the two phases, we have evaluated the purity of the steady state $P=\mathrm{Tr}(\hat\rho^2)$ systematically. The purity is close to $P=1/2$ in the symmetric region, as one would expect for a mixture of two pure states, while it increases when moving away from the phase boundary in the broken symmetry region of the phase diagram, eventually approaching the value $P=1$ for very large values of the two-photon pump $G$.

A clear picture of the two phases is obtained by simulating the mean-field dynamics towards the steady state. To this purpose, we solve Eq. (\ref{master}) numerically, taking as the initial condition a coherent state $|\alpha_0\rangle$. Fig. \ref{fig2}(e) and (f) show the trajectories on the $(\mathrm{Re}(\langle\ab\rangle),\mathrm{Im}(\langle\ab\rangle))$-plane, respectively for a symmetric and a broken-symmetry point on the phase diagram. In each case four different initial states, denoted by circles on the plots, are assumed. For the symmetric phase, all trajectories converge to a unique fixed point corresponding to the symmetric steady state. For the broken-symmetry case, depending on the initial state, the trajectories converge to two fixed points at opposite positions on the plane, again in agreement with the picture of a spontaneously broken $\mathbb{Z}_2$ symmetry. 

\begin{figure}[ht]
\includegraphics[width=0.5 \textwidth]{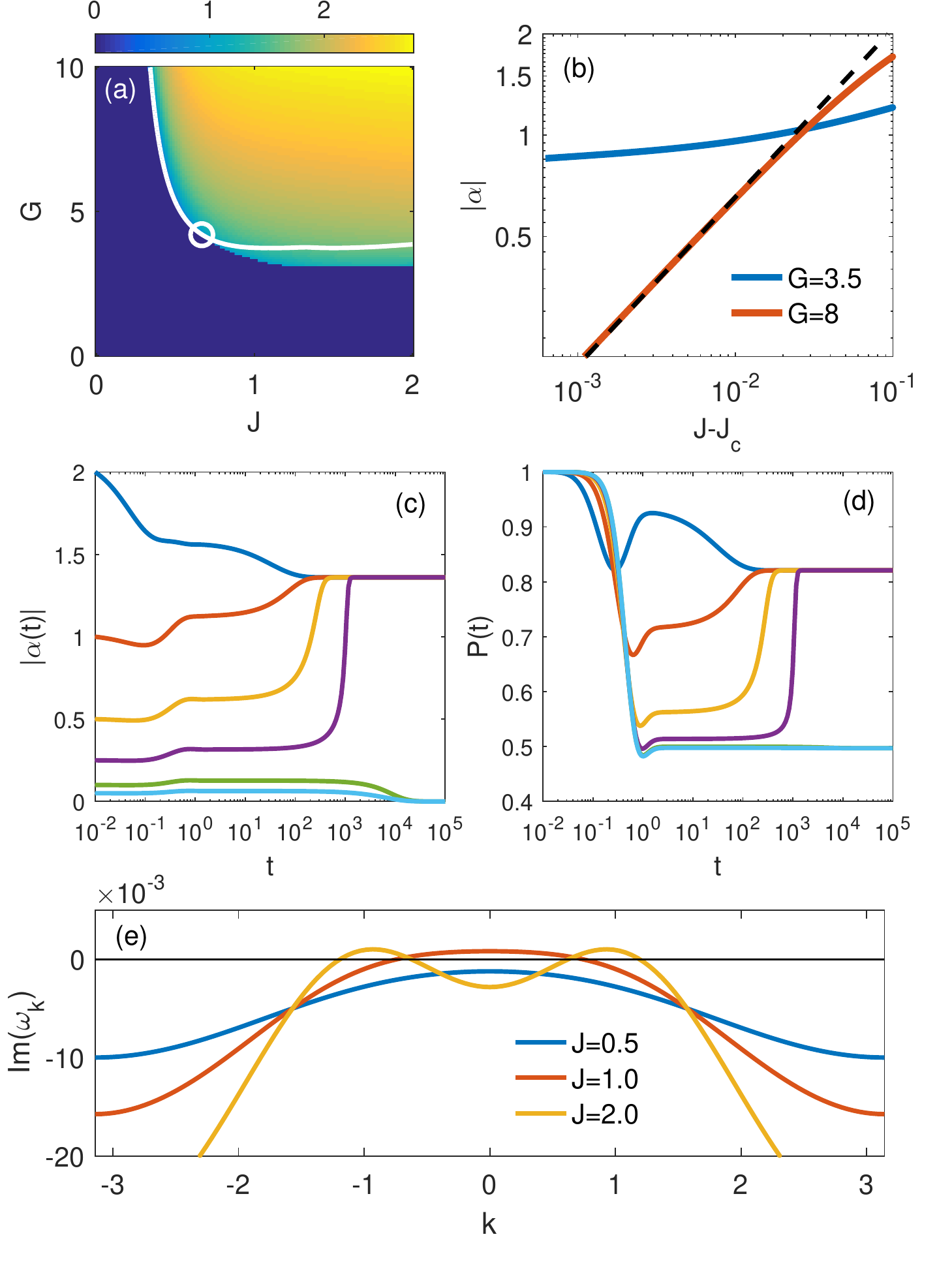}
\caption{\label{fig3} (a) White line: Contour separating the regions where $\max(\mathrm{Im}(\omega_k))|_k<0$ and $\max(\mathrm{Im}(\omega_k))|_k>0$. Color plot: the order parameter $|\langle\ab\rangle|$ computed self-consistently at steady-state. The white circle marks the onset of the bistable region of the phase diagram. (b) Double-logarithmic plot of $|\langle\ab\rangle|$ as a function of $J-J_c$ for two values of $G$. The dashed line denotes the critical behavior $|\langle\ab\rangle|\propto|J-J_c|^{1/2}$ for $G=8$, obtained by fitting the data close to the critical point $J=J_c$. (c) and (d) plots of $|\langle\ab\rangle(t)|$ and $P(t)=\mathrm{Tr}(\hat\rho^2(t))$ respectively, as a function of time, for $J=1$ and $G=3.7$. Different curves correspond to different initial coherent states $|\alpha_0\rangle$, with $\alpha_0=2.0,\,1.0,\,0.5,\,0.25,\,0.1,\,0.05$. (e) Dispersion of $\mathrm{Im}(\omega_k)$ for the least stable excitation of the symmetric steady state, as computed for three different values of $J$, and $G=4$. Parameters: $\Delta=0$, $U=\eta=1$.}
\end{figure}

We extend now the present study to the case with $\Delta=0$, which corresponds to driving the system in resonance with the bare oscillators. Fig. \ref{fig3}(a) displays the contour of the instability region and the value of the order parameter $|\langle\ab\rangle|$ computed at steady-state, as a function of $J$ and $G$. Differently from the $\Delta=-J$ case, here the region with nonzero order parameter does not completely coincide with the one where the symmetric solution is unstable. A narrow area for $J>0.67$ shows that spontaneous symmetry breaking is possible even if the symmetric solution is stable. There is in particular a ``triple point'', denoted by a circle on the plot, marking the onset of this bistable region. For values of $G\gtrsim4.2$, i.e. above the bistable region of the phase diagram, the phase boundaries defined by the two calculations coincide up to numerical accuracy, and again the phase transition displays a second-order character with critical exponent $\beta=1/2$, as shown in Fig. \ref{fig3}(b) for $G=8$. For $G<4.2$ instead, the fit in the vicinity of $J=J_c$ indicates a first order transition, as one would expect in the presence of bistability. Figs.  \ref{fig3}(c) and (d) show the time dependence of $|\langle\ab\rangle|$ and of the purity $P$, as computed for one point lying in the bistable region of the phase diagram. Different curves correspond to different initial values of $\alpha_0$. Depending on the displacement of the initial coherent state, the dynamics converges either to the symmetric or to the broken-symmetry phase. In the first case, the purity reaches a value close to $0.5$ as expected, while in the second case it approaches a higher value, as discussed above. For the initial state with $\alpha_0=0.25$, which is closest to the transition between the two types of dynamics, we notice a long-time metastable transient which lasts up to $10^3$ inverse lifetimes and is reminiscent of the metastable states discussed in the literature for the single-site Kerr model \cite{Minganti2016}. We find a similar behavior, with a long-lived metastable transient, when approaching the second-order phase boundary (not shown). This feature can be interpreted as a manifestation of the critical slowing down typically present in the vicinity of a phase boundary.

The bistability observed here for $\Delta=0$ -- i.e. when driving {\em above} the bottom of the band -- is reminiscent of the bistable behavior of a Kerr oscillator when driven with a positively detuned laser frequency \cite{Drummond1980}. While in the linearly driven case the bistability is at the origin of the critical behavior \cite{Biondi2016,Foss-Feig2017,LeBoite2014,LeBoite2013,Wilson2016}, here it competes with the genuine second-order phase transition enabled by the two-photon driving. In a description of the system beyond the mean-field approximation, we expect this $k=0$ bistability to compete with the instability associated to the optical parametric oscillator, where two $k=0$ pump photons injected above the band bottom, scatter resonantly to a $\pm k$ pair of states \cite{Carusotto2013}. Evidence is again obtained from the stability analysis of the symmetric state. In Fig. \ref{fig3}(e), the dispersion of the excitation with the largest values of $\mathrm{Im}(\omega_k)$ is shown for $G=4$ and $J=0.5,\,1.0,\,2.0$, as computed for $\Delta=0$. The instability can emerge at opposite values of $k\ne0$, as for the case with $J=2.0$, indicating the onset of the parametric oscillation. In this case, we expect a broken-symmetry phase characterized by more exotic correlation patterns, similarly to what was recently predicted for the driven-dissipative Rabi-Hubbard model \cite{Schiro2016}. It should be noted that in the $\Delta=-J$ case the instability always emerges at $k=0$, further highlighting the genuine second-order character of this phase transition.

\section{Discussion}

The present mean-field analysis provides a first hint that the phase transition associated to the $\mathbb{Z}_2$ symmetry should actually occur in an array of coupled quadratically driven Kerr resonators, and would represent the quantum analog of a classical Ising simulator, that is realized in the limit of large two-photon driving field $G$ \cite{Marandi2014,Wang2013}. It is important to highlight here that the $\mathbb{Z}_2$-manifold that characterizes the phase transition is generated by the very specific driven-dissipative protocol \cite{Leghtas2014}, which in a single resonator results in a degenerate pair of cat states protected from the environment. In an array of resonators governed by the Hamiltonian (\ref{H}), such a $\mathbb{Z}_2$-manifold is still present and lies in the excited region of the spectrum, far from the ground state, as suggested already by a study of two coupled resonators \cite{Nigg2017}. Hence, the phase transition described here is a purely nonequilibrium phenomenon enforced by the driven-dissipative nature of the system -- similarly to the case of the incoherently-driven Bose-Hubbard model \cite{Biella2017} -- and represents an experimentally viable example of a genuinely dissipative phase transition \cite{Kessler2012}. This marks a substantial difference with respect to other driven-dissipative systems where a phase transition associated to a $\mathbb{Z}_2$ symmetry breaking has been investigated -- such as the Rabi-Hubbard lattice \cite{Schiro2012,Schiro2016} and the Dicke model \cite{Baumann2010,Baumann2011,Brennecke2013,Dimer2007} -- in that the transition in those cases is a ground-state property inherited from the quantum phase transition of the corresponding Hamiltonian system, while the driven-dissipative nature of the system results in peculiar features such as modified critical exponents \cite{Brennecke2013} or exotic attractors \cite{Schiro2016}. 

An experimental platform that would naturally behave according to the model studied here, is that of superconducting circuits, where two-photon driving was achieved in a scheme where two microwave resonators are coupled through a Josephson junction \cite{Leghtas2014}. For this system, photonic Schr\"odinger's cat states were experimentally characterized in full agreement with the quadratically driven Kerr model. The extension to an array of coupled resonators is also possible, as linear coupling between superconducting microwave resonators has recently been demonstrated in several experiments \cite{Eichler2014,Fitzpatrick2017}. In the optical domain, polaritons in semiconductor microcavities are an alternative promising system for a physical implementation of the present model. Polaritons are naturally endowed with a Kerr nonlinearity \cite{Carusotto2013}, arrays of coupled polariton micropillars are now routinely fabricated in several geometries \cite{Baboux2016,Jacqmin2014}, and two-photon driving of polaritons was recently demonstrated \cite{Kavokin2012,Lemenager2014,Steger2015}.

In view of a clear experimental characterization of the phase transition, superconducting circuits represent the election system, thanks to the possibility to carry out Wigner function tomography \cite{Leghtas2014}. For systems in the optical spectral range on the other hand, a signature of the phase transition should emerge from the measurement of the second order correlation function of the emitted light, as was recently demonstrated on a polariton system in the presence of a first order phase transition \cite{Fink2017}.

\section{Conclusions}

We have investigated the model of a driven-dissipative array of coupled Kerr resonators in the presence of two-photon driving. The mean-field analysis shows a clear signature of an Ising-like phase transition associated to the spontaneous breaking of a $\mathbb{Z}_2$ symmetry. Our finding provides a simple answer to the question whether a coherently driven Bose-Hubbard system may still display a critical behavior associated to a spontaneous symmetry breaking. Here, contrarily to the case with one-photon driving, the original $U(1)$ symmetry of the Bose-Hubbard model is not completely lifted by the conherent driving field, and a $\mathbb{Z}_2$ symmetry is left in the system.   The next step would consist in an analysis beyond mean-field. To this purpose, methods may include cluster mean-field \cite{Jin2016}, truncated correlation hierarchy schemes \cite{Casteels2016}, Langevin Monte Carlo approaches deriving from quasi-probability distributions \cite{Deuar2002,Vogel1989}, and ultimately large scale numerical schemes for small lattices \cite{Finazzi2015}. The present study could be extended to the transverse Ising model in the presence of an additional linear driving term or -- more interestingly -- to interacting spin models when in presence of cross-Kerr nonlinearity, and to more exotic models with the introduction of $N$-photon driving terms \cite{Devoret2014}. Finally, given the great promise held by the quadratically driven Kerr system as a building block of a photonic quantum information platform \cite{Goto2016,Goto2016a,Nigg2017,Puri2017a,Puri2017}, it is also important to further investigate the role of the critical behavior studied here, in the context of these applications. 

\begin{acknowledgements}
I am indebted to Eduardo Mascarenhas for several insightful discussions and precious suggestions. I acknowledge enlightening discussions with Gianni Blatter, Matteo Biondi, Cristiano Ciuti, Nicola Bartolo, and Fabrizio Minganti.
\end{acknowledgements}

\bibliographystyle{apsrev4-1}

\end{document}